\documentclass[%
 reprint,
 superscriptaddress,
 amsmath,amssymb,
floatfix,
]{revtex4-2}

\usepackage{siunitx}
\usepackage{cases}

\usepackage{graphicx}
\usepackage{dcolumn}
\usepackage{bm}
\usepackage{amsthm}
\usepackage{comment}
\usepackage{hyperref}
\hypersetup{
    colorlinks=true,
    linkcolor=blue,
    citecolor=blue,
    filecolor=blue,
    urlcolor=blue,
}

\begin{document}

\title{ Magnetically controlled double-twist director configuration of lyotropic chromonic liquid crystals in cylinders: Energetics, topological defects, and instability}

\author{Junghoon Lee}
\affiliation{
 Department of Physics, Ulsan National Institute of Science and Technology, Ulsan, Republic of Korea
}
\author{Joonwoo Jeong}
\email{jjeong@unist.ac.kr}
\affiliation{
 Department of Physics, Ulsan National Institute of Science and Technology, Ulsan, Republic of Korea
}

\date{\today}

\begin{abstract}
We study experimentally how the double-twist (DT) configuration of cylindrically confined lyotropic chromonic liquid crystals (LCLCs) responds to axial magnetic fields. Our director field model unveils the energetics behind the magnetic field-induced transition in the twist profile of the DT configuration. Additionally, we catalog three different types of topological defects---residing between the DT domains of opposite handedness---before and after the field application, and propose a new director field model for the defect with a ring disclination. Lastly, we report a symmetry-breaking instability occurring when the field strength exceeds a critical value, suggesting an eccentric DT director field model that reproduces a helix-like optical texture. Our systematic investigation not only enhances our understanding of LCLC energetics but also provides potential for precise control over DT configurations.
\end{abstract}

\maketitle


\section{Introduction}

Liquid crystals (LCs), as a prime example of soft matter, have been the subject of extensive study due to their significant response functions \cite{deGennesSoftMat}. This research has enhanced our understanding of these partially ordered condensed matter and has paved the way for various applications, such as information displays \cite{Chandrasekhar1992, Gennes1993, Lavrentovich2020}. Specifically, LCs are characterized by their responses to external stimuli such as temperature, light, and mechanical stress \cite{Sengupta2014, Sagues2015, Smalyukh2019, Ignes2025}. In particular, LCs under electric or magnetic fields often show the Fr\'eedericksz transition, which is used to measure the physical properties of LCs, including elastic moduli, viscosities, and anchoring coefficients \cite{Zhang2003, Zhou2012PRL}. This transition also plays a crucial role in developing reversible switching mechanisms for applications in photonics, sensors, and actuators \cite{Coles2010,Munna2019,Gruzdenko2022}.

In contrast to thermotropic LCs, the investigation of lyotropic LCs' responses to electric and magnetic fields has been relatively limited \cite{Lekkerkerker2013, Kadar2021, Ferreira2022, PDavidson2024}. The compositional complexity of lyotropic LCs, such as the presence of ions in an aqueous solvent and other components that are incompatible with the fields, can complicate isolated studies of the interactions between LC phases and external fields \cite{Kang2015}. Additionally, experimental challenges such as high viscosity and small dielectric or magnetic anisotropy, which often lead to long relaxation times, can hinder observations in the laboratory \cite{Giesselmann2024, Mezzenga2025}.

In the same vein, there are only a few reports on lyotropic chromonic liquid crystals (LCLCs) under electric and magnetic fields, despite their significant potential highlighted in recent years \cite{Lee1982, Hui1985, Ignes2020Cryst, Boule2020, Li2022}. LCLCs, partially ordered aggregates in which planar molecules are stacked via non-covalent interactions in water, often exhibit unusual LC properties: exceptionally high order parameters, elastic anisotropy, biocompatibility, and strong dependence on solvent and additives \cite{Lydon2010Review, Lydon2011Review,TortoraPNAS2011,Zhou2012PRL,ZhouPNAS2014,JeongPNAS2014,JeongPNAS2015,Davidson2015,Nayani2015,Arman2018,Lee2019,Eun2019,Yang2022,Kim2024,Cheon2025}. As a result, LCLCs have been adopted as templates for creating highly ordered structures \cite{TamChang2008, Asdonk2016, Bosire2021}, as stimuli-responsive platforms for living organisms \cite{Carlton2013, Mushenheim2015, Peng2016, Shaban2021}, and as model systems for studying the topology and symmetry breaking of partially ordered matter \cite{Dietrich2021, Eun2021}. Beyond these passive observations, active control of LCLC configurations using magnetic fields \cite{Ignes2020Cryst, Boule2020} could enhance our understanding of these materials and expand their range of applications. 

In this study, we experimentally investigate the energetics, topological defects, and instability of cylindrically confined nematic LCLCs subjected to a sub-Tesla axial magnetic field implemented by a Halbach array of permanent magnets \cite{Ignes2016PNAS, Ignes2020Cryst}. In terms of elastic free energy, taking into account LCLC's large saddle-splay elastic modulus and negative magnetic anisotropy, we elucidate how its chiral director configuration responds to the applied magnetic field. Additionally, we report multiple types of topological defects under a magnetic field, as well as instabilities observed in high-field regimes. Our findings suggest that magnetic field-controlled LCLCs warrant further investigation.

\section{Results and Discussion}

\begin{figure}[t]
\centering
  \includegraphics{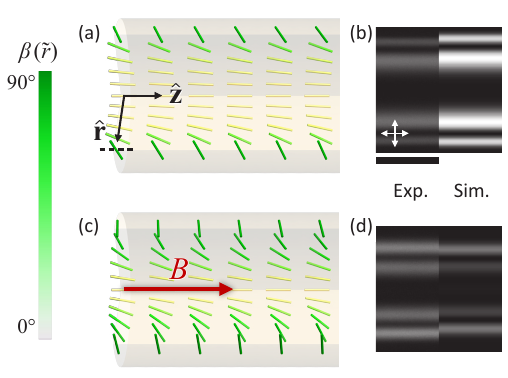}  
  \caption{\label{fig:schematics} 
Double-twist (DT) configuration under an axial magnetic field. (a) The director field of the DT configuration at $B = 0$. The angle $\beta$ between the nematic director (a colored rod) and the capillary axis is indicated by the color. (b) The polarized optical microscopy (POM) image of the DT configuration (left) and the corresponding Jones-calculus simulated optical texture (right). We experimentally observe the DT configuration of 14~wt\% nematic DSCG confined in a capillary with a 100-\si{\micro\meter} diameter, using crossed polarizers depicted as white double arrows. The scale bar is 50-\si{\micro\meter}. The DT configuration at $B = 0.33~\si{\tesla}$ is shown in (c), and the experimental image and the corresponding simulated textures in (d).
}
\end{figure}

We observe the response of nematic LCLC's double-twist (DT) configurations in a cylinder to a magnetic field applied along the cylindrical axis. Figure~\ref{fig:schematics}(a) and (c) show the director field of the well-known DT configuration of 14~wt\% nematic DSCG in a cylindrical capillary \cite{Crawford1993,Davidson2015,Nayani2015,Urbanski2017,Eun2019}, an experimentally observed polarized optical microscopy (POM) texture, and the corresponding Jones calculus-simulated texture. See \hyperref[sec:mnm]{Materials and Methods} for the details. As shown in Fig.~\ref{fig:schematics}(d), upon applying the axial magnetic field, the optical texture changes in terms of the positions and relative brightness of the stripes, indicating alterations in the nematic director configuration. 

We numerically find the elastic free energy-minimizing director configuration to elucidate this observed change. With the nematic director field of the DT configuration \cite{Eun2019}, $\mathbf{n}=\sin{\beta}\ \mathbf{\hat{r}}+\cos{\beta}\ \mathbf{\hat{z}}$, we write the elastic free energy $F$ in the cylindrical coordinate system $(r, \theta, z)$.
\begin{multline}\label{eq:felastic}
        F=\int dV \left[  \frac{1}{2}K_1(\nabla\cdot\mathbf{n})^2+\frac{1}{2}K_2(\mathbf{n}\cdot\nabla\times\mathbf{n})^2\right. 
        \\+\frac{1}{2}K_3(\mathbf{n}\times\nabla\times\mathbf{n})^2
        -\frac{1}{2}K_{24}\nabla\cdot(\mathbf{n}\times\nabla\times\mathbf{n}+\mathbf{n}\nabla\cdot\mathbf{n})
        \\\left.-\frac{1}{2}\mu_0^{-1}\chi_a(\mathbf{n}\cdot\mathbf{B})^2
        \right],
\end{multline}
where $K_1$, $K_2$, $K_3$, and $K_{24}$ are splay, twist, bend, and saddle-splay elastic moduli, respectively. In the $F$'s last term regarding the magnetic field $\mathbf{B}$, $\mu_0$ is the permeability of free space, and $\chi_a$ is the magnetic anisotropy of LCLC. Note that the surface anchoring energy does not play a role because the capillary wall imposes the degenerate planar anchoring condition \cite{Davidson2015}. Applying the calculus of variation extremizing $F$ with the magnetic field $\mathbf{B}=B\mathbf{\hat{z}}$, we derive the Euler-Lagrange differential equation Eq.~(\ref{eq:deq}) and corresponding boundary conditions Eq.~(\ref{eq:bcs}).
\begin{equation}\label{eq:deq}
    \begin{split}
        \frac{k_2\sin{4\beta}}{2}+&4\sin^3{\beta}\cos{\beta}-2k_2\tilde{r}\beta'\\&-2k_2\tilde{r}^2\beta''+\tilde{r}^2b\sin{2\beta}=0,
    \end{split}
\end{equation}
\begin{equation}\label{eq:bcs}
    \begin{split}
        \beta(\tilde{r}=0)=0,\ \beta'(\tilde{r}=1)=\frac{k_{24}-k_2}{2k_2}\sin{2\beta(\tilde{r}=1)}
    \end{split}
\end{equation}
Here we introduce the dimensionless variables: $\tilde{r}=r/R$, $k_2=K_2/K_3$, $k_{24}=K_{24}/K_3$ and $b=R^2B^2\chi_a/(\mu_0 K_3)$, where $R$ denotes the capillary radius. 
The variable $b$ regarding the magnetic field depends on the capillary size $R$, which means that the field effect should be size-dependent, while the effect of the saddle-splay elasticity is size-independent \cite{Davidson2015}.

\begin{figure}[t]
\centering
  \includegraphics{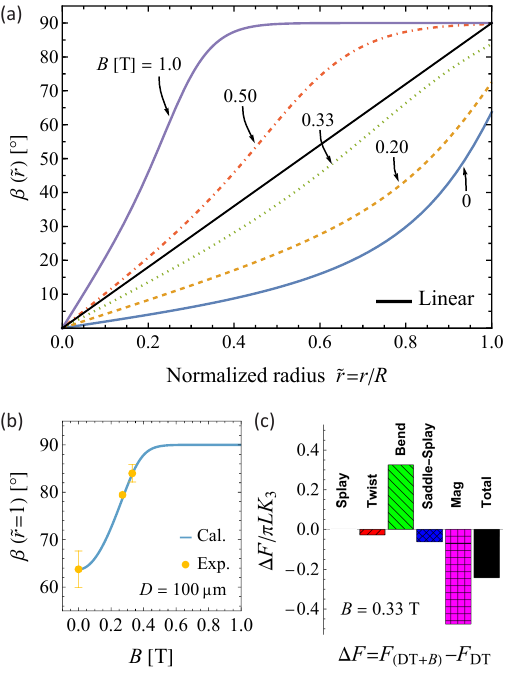}
  \caption{\label{fig:betaProfile} 
Numerically calculated DT configurations under axial $B$ fields and their comparison with experimental data. (a) The twist angle $\beta$ profile along the normalized radius $\tilde{r}=r/R$ according to the strengths of the axial magnetic field $B$. The material parameters used for all calculations in Fig.~\ref{fig:betaProfile} are from a 14~wt\% DSCG in a capillary with a 100-\si{\micro\meter} diameter. (b) The twist angle at the capillary wall, $\beta_1$, according to $B$. The solid line is from the numerical calculation, and the symbols correspond to the experimental measurements. The error bars represent the standard deviations from five independent experimental replicates. The data point without an error bar at $B= 0.27~\si{\tesla}$ is from a single measurement. (c) Free energy differences between the zero-field and 0.33~\si{\tesla}-applied DT configurations. The energy differences per unit length ($\Delta F/L$) of each deformation mode and their sum are non-dimensionalized by the bend elastic modulus $K_3$.
}
\end{figure}

Figure~\ref{fig:betaProfile}(a) presents the energy-minimizing twist profiles $\beta(\tilde{r})$ of the DT configuration $\mathbf{n}=\sin{\beta}\ \mathbf{\hat{r}}+\cos{\beta}\ \mathbf{\hat{z}}$ according to the magnitude of the axial magnetic field $\mathbf{B}$, exhibiting a convex-to-concave transition. We adopt $D=2R=100~\si{\micro\meter}$, $k_2=1/30$, $k_{24}=1/2$ \cite{Arman2018,Eun2019}, $\chi_a=-3.5\times10^{-7}$ \cite{Zhou2017Book} for 14~wt\% DSCG. Deviating from the analytically solvable $\beta(\tilde{r})$ at $B=0$, the profile changes its shape from a convex curve to a concave one, and the twist angle at the capillary wall,  $\beta_1=\beta\left(\tilde{r}=1\right)$, increases as $B$ increases, as shown in Fig.~\ref{fig:betaProfile}(b). A representative calculation in Fig.~\ref{fig:betaProfile}(c) illustrates the energetics behind this transition: a competition between magnetic alignment and bend penalty. Specifically, because of the negative magnetic anisotropy, the magnetic energy decreases as the directors align perpendicular to the field, making $\beta$ close to 90\si{\degree} upon application of the axial magnetic field. On the other hand, $\beta$ close to 90\si{\degree} increases the bend energy, especially at the core region with a small radius of curvature. This competition mainly determines the energy minimum at a given magnetic field strength, where the decrease in the magnetic energy surpasses the increase in the bend energy, thereby minimizing the total energy, as shown in Fig.~\ref{fig:betaProfile}(c). The convex-to-concave transition is also captured in our Landau-de Gennes calculation using open-Qmin in the $\mathit{Q}$-tensor formalism \cite{Ravnik2009LdG,Zhou2017Cryo,openQmin,Long2021, Ettinger2022, Ziga2024}. (See Supplemental Material)

Experimentally estimated $\beta_1$ agrees with the theoretically calculated values, as shown in Fig.~\ref{fig:betaProfile}(b). We estimate the twist angle at the capillary wall $\beta_1$ from POM images and find that it increases as we increase the magnetic field. For instance, 14~wt\% DSCG confined in 100-\si{\micro\meter} capillaries without the magnetic field exhibit $\beta_1 = 64\pm 4 \si{\degree}$, and $\beta_1 = 84\pm 2 \si{\degree}$ after applying 0.33-\si{\tesla} axial magnetic field. Since LCLC's director relaxation in the magnetic field takes hours, we ensure the director configuration reaches a steady state before estimating $\beta_1$ (See Supplemental Material). As shown in Fig.~\ref{fig:betaProfile}(b), experimentally estimated $\beta_1$ indicated by a symbol increases as $B$ increases, lying on the theoretically predicted curve. We also find that $B$ affects the birefringence $\Delta n$, and a hysteresis exists; the $\beta_1$ and $\Delta n$ do not recover the zero-field value after removing the magnetic field. Additionally, we inspect another well-studied LCLC, sunset yellow FCF (SSY), and find that it similarly responds to the magnetic field. The data regarding the hysterisis and SSY is included in the Supplemental Material and deserves further investigation. 

\begin{figure*}[t]
\centering
  \includegraphics{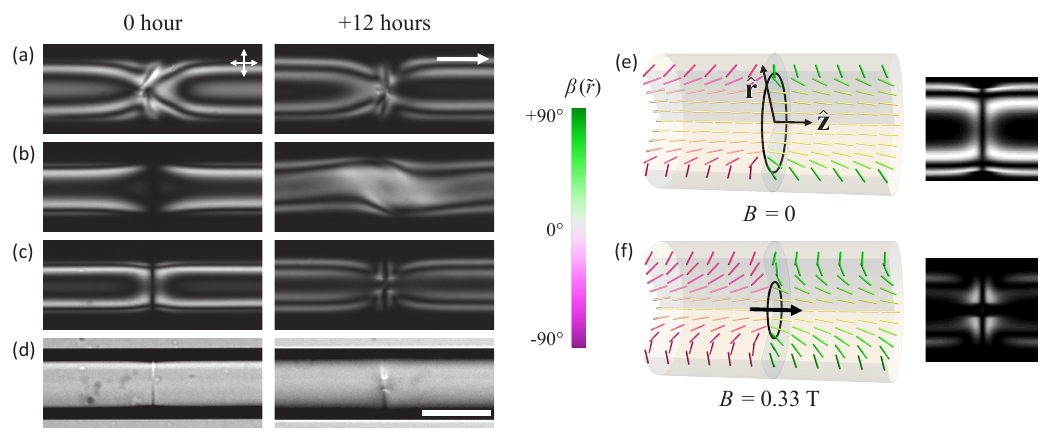}
  \caption{\label{fig:defect} 
  Three different types of defects between DT domains and their responses to the axial magnetic field. (a) Experimental POM image of a point defect before and after applying the axial magnetic fields of 0.33~\si{\tesla}. The white double arrows represent the pass axes of polarizers, and the thick right arrow indicates the direction of the magnetic field. (b) A domain-wall defect. (c) A ring disclination. (d) The bright-field image of the ring disclination.
  (e-f) Our director model for the ring disclination. The colored rods represent the nematic director and its $\beta$ value. The ring illustrated at the center is a revolution of $s=+1/2$ disclination, and the black arrow in (f) indicates the direction of the magnetic field. The ring disclination shrinks as the magnetic field strength increases: (e) $\tilde{r}=0.72$ at $B=0$ and (f) $\tilde{r}=0.41$ at $B=0.33~\si{\tesla}$. 
}
\end{figure*}

We observe that the magnetic field also affects the structure of defects at the interface between two DT domains of different twist handednesses. Unlike the DT configuration of SSY, which only exhibits a point defect \cite{Davidson2015}, three different types of defects exist in the DT configuration of DSCG \cite{Madina2022}: a point defect in Fig.~\ref{fig:defect}(a), a domain wall in Fig.~\ref{fig:defect}(b), and an unidentified singular defect in Fig.~\ref{fig:defect}(c). The point defect in Fig.~\ref{fig:defect}(a) and the defect after applying the axial magnetic field are both asymmetric, in contrast to SSY's symmetric point defect \cite{Davidson2015}. The domain wall predicted by Davidson et al. \cite{Davidson2015} is observed in DSCG \cite{Nayani2015, Madina2022}, but its symmetry is broken upon the application of the magnetic field, as shown in Fig.~\ref{fig:defect}(b). To explain these asymmetric structures, modifying the defect models in Davidson \textit{et al.} is required in future work.

We propose that the defect in Fig.~\ref{fig:defect}(c) is a ring of $s=+1/2$ disclination \cite{Madina2022}, and the ring shrinks upon the application of the axial magnetic field. The bright-field image shown in Fig.~\ref{fig:defect}(d) hints that the defect is singular before and after the application of the axial magnetic field, and the diameter of the ring contracts when subjected to the magnetic field. Our director field model (Eq.~(\ref{eq:ring})) is sketched in Fig.~\ref{fig:defect}(e) and (f) rationalizes this shrinkage.
  \begin{equation}\label{eq:ring}
      \begin{split}
          \beta_{\mathrm{ring}}(\tilde{r},\tilde{z})=\frac{\pi e^{-|4\tilde{z}|}}{2+2e^{-(\tilde{r}-\tilde{r}_0)/\tilde{w}_{\mathrm{ring}}}}+(1-e^{-|4\tilde{z}|})\beta(\tilde{r})
      \end{split}
  \end{equation}
Here, all length variables are nondimensionalized by the capillary radius $R$. At a given field strength, we minimize the elastic free energy with respect to the nondimensionalized ring radius $\tilde{r}_0$ while fixing $\tilde{w}_{\mathrm{ring}}=0.01$. The Jones calculus-simulated optical textures that best match the experimental POM images are shown in Fig.~\ref{fig:defect}. As the field strength increases, the radius of the energy-minimizing ring decreases. In the circular cross-section of the cylinder at $\tilde{z}=0$ containing the disclination ring, the directors inside the ring are parallel to the axial magnetic field, and they are energetically unfavorable because of LCLC's negative magnetic anisotropy. On the contrary, the directors outside of the ring, which are perpendicular to the field, can lower the magnetic free energy. Thus, upon exposure to the magnetic field, the ring may shrink to reduce the energy. 

\begin{figure*}[t]
\centering
  \includegraphics{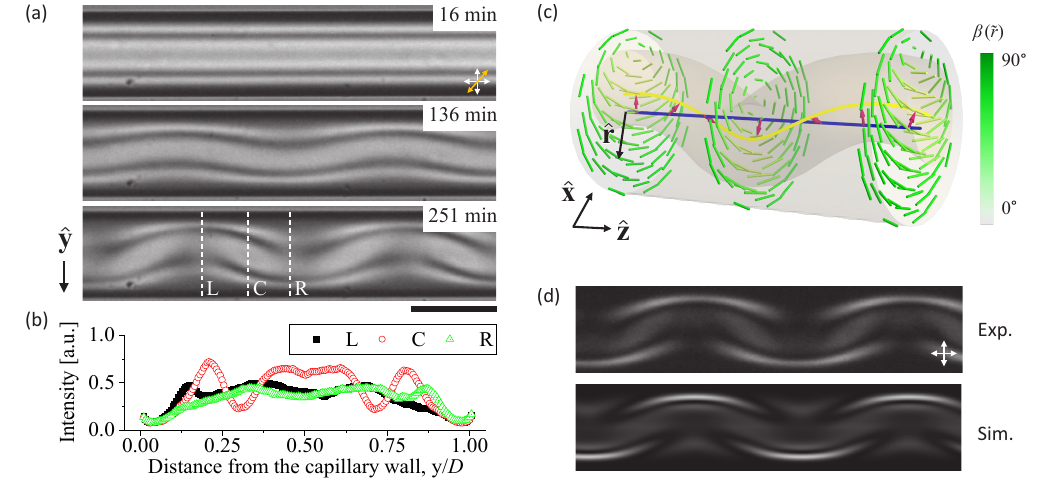}
  \caption{\label{fig:wavy}
  Magnetic-field induced instabilities in the DT configuration. (a) After applying $B = 0.33~\si{\tesla}$, a wavy optical texture develops. The text displayed at the top right corner of each image indicates the time elapsed after applying the magnetic field. The white double arrows indicate the pass axes of the crossed polarizers, and the yellow diagonal arrow is the fast axis of the full-waveplate ($\Delta\lambda=530~\si{\nano\meter}$). The scale bar is 100~\si{\micro\meter}. (b) Grayscale intensities of the wavy texture show its periodicity. Measured intensities along dashed lines at the left (L) and right (R) sides of the wave's node are asymmetric with respect to the capillary axis, whereas the intensity at the node (C) shows symmetry. Distance from the upper capillary wall $y$ is normalized by the capillary diameter $D=2R$.
  (c) Our director model for eccentric double-twist configuration in a cylinder. The colored rods represent the nematic director and its $\beta$ value. The blue straight line parallel to $\hat{z}$ indicates the capillary axis, while the yellow helical line shows the core of the double-twist configuration. Purple arrows indicate the deviation of the core from the capillary axis. 
  (d) Comparison of an experimental POM texture (top) and a Jones-calculus simulated optical texture of our model (bottom).
  }
\end{figure*}

Lastly, a wavy optical texture shown in Fig.~\ref{fig:wavy}(a) appears when the strength of the axial magnetic field exceeds a critical value. The critical field strengths $B_c$ are approximately \qtylist{0.34;0.31}{\tesla} for capillaries with diameters of \qtylist{100;150}{\micro\meter}, respectively. In contrast to the axisymmetric optical textures with straight stripe patterns, the wavy texture is periodic with the wavelength $\sim 5R$ in 100-\si{\micro\meter} capillaries under $B=0.38~\si{\tesla}$, where $R$ is the radius of the cylinder. The time evolution of the texture for 4~hours, until it reaches a steady state, is shown in Fig.~\ref{fig:wavy}(a).

We hypothesize that an eccentric double-twist (EDT) configuration is responsible for the wavy texture. The POM texture shown at the bottom row of Fig.~\ref{fig:wavy}(a) indicates that the director configuration is no longer radially symmetric. Specifically, the intensity profiles along the diameter periodically change as $z$ changes, oscillating between asymmetric and symmetric ones. We presume that a DT-like configuration has an off-centered core at the $x$\nobreakdash--$y$ plane, and the position of the core oscillates as $z$ changes. Note that we exclude a scenario in which the core oscillates within a particular plane, e.g., $y=0$, because all observed capillaries exhibit the same texture despite the absence of an oscillation-plane selection mechanism. Given that there is no preferred oscillation plane, we propose that the off-centered core has a helical structure about the capillary axis, $z$-axis. This eccentric double-twist configuration has been adopted to explain similar instabilities observed in cholesteric liquid crystals \cite{Kitzerow1996,Zumer1999,Viviana2023}. 

We present an empirical director field model for the EDT configuration that is modified from the DT configuration in three steps. 
First, to create the off-centered core, we translate the DT configuration along the $x$-axis according to a nondimensionalized offset $b(\tilde{r})$ as a function of $\tilde{r}$.
Second, to create the helical structure, we rotate the directors at each $\tilde{z}$-plane about the $z$-axis by $\theta(\tilde{z}) = 2 \pi \tilde{z} /\tilde{\lambda}$, where $\tilde{\lambda}$ is a nondimensionalized helical pitch.
Lastly, we rotate each director in the translated and rotated configuration about the $x$-axis by an angle function $t(\tilde{r})$, lowering the elastic free energy. 
We empirically adopt $b(\tilde{r}) = b_0 (\tilde{r}-1)$ and $t(\tilde{r}) = \theta_0 (1-\tilde{r}^3)$ and connect $\theta_0$ with the pitch $\tilde{\lambda}$ via $\theta_0 = 2\pi \tilde{b}_0/\tilde{\lambda}$. 
The director field of this ansatz, having $\tilde{b}_0=0.3$ and $\tilde{\lambda}=5$, is Fig.~\ref{fig:wavy}(c), and Fig.~\ref{fig:wavy}(d) compares experimental and simulated optical textures. Given the promising similarity, the detailed energetics and wavelength selection mechanism of our director field model warrant further study.

\section{Materials and Methods}\label{sec:mnm}
\subsection{Sample preparation}
The chromonic molecules used for LCLC preparation, disodium cromomglycate (DSCG) and sunset yellow FCF (SSY), were purchased from Sigma-Aldrich and dissolved into deionized water (18.2~\si{\mega\ohm\cm}). While we used the DSCG of 99.7\% purity as received, the SSY of $\geq90\%$ purity was purified before use \cite{Horowitz2005,Eun2020}. Nematic LCLCs, 14.0~wt\% DSCG and 30.0~wt\% SSY, were injected into cylindrical capillaries (CV1017 and CV1525, Vitrocom) with two different inner diameters: 100 and 150~\si{\micro\meter} with $\pm10\%$ tolerance. The epoxy-sealed capillaries were sandwiched between a glass substrate and a glass coverslip and immersed in refractive index-matching oil ($n=1.4740$ at 25\si{\degreeCelsius}, Cargille Labs). Before observation, the samples were stored at 65\si{\degreeCelsius} for 1 hour, cooled down to 22\si{\degreeCelsius} at the rate of 20\si[per-mode = symbol]{\degreeCelsius\per\minute}, then relaxed at the same temperature for 4 hours on a temperature controller (LTS120/T96, Linkam Scientific Instruments). Lastly, the samples were placed into a custom-built holder for magnetic field application and maintained at 22\si{\degreeCelsius} using a PID-controlled heater (HT10K, Thorlabs) throughout the experiments.

\subsection{Magnetic field application and polarized optical microscopy}
We built a custom sample holder for use in polarized optical microscopy under a uniform magnetic field. To apply a uniform in-plane B field to capillaries, we constructed a $k=2$-type Halbach array consisting of eight $\qtyproduct{1x1x1}{\arcsecond}$ permanent magnets (N52 grade, KJ Magnetics, Inc.) adopting the array design in the reference \cite{Ignes2016PNAS, Ignes2020Cryst}. Our cylindrical Halbach array achieved a field strength up to $B=0.38~\si{\tesla}$ at the center of the cylinder. We calibrated the magnetic field strength as a function of the position along the cylindrical axis using a gaussmeter (TM-197, Tenmars Electronics) and controlled the sample positions to apply different field strengths. Note that we cut the lengths of the capillaries to be smaller than 25~\si{\milli\meter} to make the applied field as uniform as possible.

The custom holder was attached to an upright microscope (BX40, Olympus) equipped with a color CMOS camera (DigiRetina 16, Tucsen Photonics) and quasi-monochromatic illumination with a bandpass filter having $\lambda = 660~\si{\nano\meter}$ and $\textrm{FWHM} =10~\si{\nano\meter}$ (FBH660-10, Thorlabs). A polarizer, an analyzer, and a full-wave plate (U-TP530, Olympus, $\Delta\lambda = 530~\si{\nano\meter}$) in front of the analyzer were placed accordingly.

\subsection{Comparison between experimental POM images and Jones calculus-simulated optical textures}
To validate our calculation, we compared experimental POM images with Jones-calculus simulated optical texture. We measured the grayscale intensities of cylinders' central regions ($\sim 3\%$ of diameter) from the experimental POM images, varying the analyzer angle while keeping the polarizer angle fixed. This intensity profile, as a function of the analyzer angle, can be compared to the transmittance profile generated from Jones calculus-simulated optical textures \cite{Davidson2015,Eun2019}. Specifically, for a given capillary diameter matching the experimentally measured one, we found the best-matching intensity profile that minimized the root-mean-square error, varying the birefringence $\Delta n$ and the twist angle $\beta_1$ at the capillary wall. The estimated $\Delta n$s are comparable to the previously reported measurements \cite{Nastishin2005}, and the simulated textures showed good agreement with the POM images, as shown in Fig.~\ref{fig:schematics}(b) and (d).

\section{Conclusion}

In summary, we report on the response of the DT configuration of cylindrically confined LCLCs and the topological defects therein to axial magnetic fields. Experimental observations agree with our director field model, which minimizes the Oseen-Frank elastic free energy considering the LCLC's negative magnetic anisotropy. The twist profile along the radius undergoes a transition from concave to convex shape with increasing magnetic field strength. Moreover, when the field strength exceeds a threshold value, the radial symmetry of the DT configuration breaks, presumably exhibiting an eccentric DT configuration with a helical core. The energetics of this instability and the topological defects of three different types require further investigation in future work. Additionally, considering that LCLC's responses to electric fields are complicated due to its aqueous and ionic nature, we envision that this sub-Tesla magnetic-field application---achievable by a Halbach array of permanent magnets on a tabletop---will deepen our understanding of LCLC energetics, beyond the DT configurations in cylinders. For instance, the magnetic field can be utilized to study LCLC's polar anchoring strengths \cite{Jeong2014, Asdonk2017}, chiral nematic LCLCs under confinement \cite{Eun2021, Pisljar2022, Chernyshuk2023}, and even theories beyond the Oseen-Frank description \cite{Paparini2024, Ciuchi2024}.

\begin{acknowledgments}
The authors gratefully acknowledge financial support from the Korean National Research Foundation through Grants No. NRF-2023K2A9A2A23000311 and RS-2024-00345749. We also thank Prof. Jaeup U. Kim's group for providing their computing resources and Dr. Sung-Jo Kim for the fruitful discussion.

\end{acknowledgments}

%

\end{document}



\title{Supplemental Materials for ``Magnetically controlled double-twist director configuration of lyotropic chromonic liquid crystals in cylinders: Energetics, instability, and topological defects''}

\author{Junghoon Lee}
\affiliation{
 Department of Physics, Ulsan National Institute of Science and Technology, Ulsan, Republic of Korea
}
\author{Joonwoo Jeong}
\email{jjeong@unist.ac.kr}
\affiliation{
 Department of Physics, Ulsan National Institute of Science and Technology, Ulsan, Republic of Korea
}

\date{\today}

\maketitle

\section{AFM investigation of the inner surface topography of a glass capillary}
  \begin{figure}[ht]
  \centering
  \includegraphics{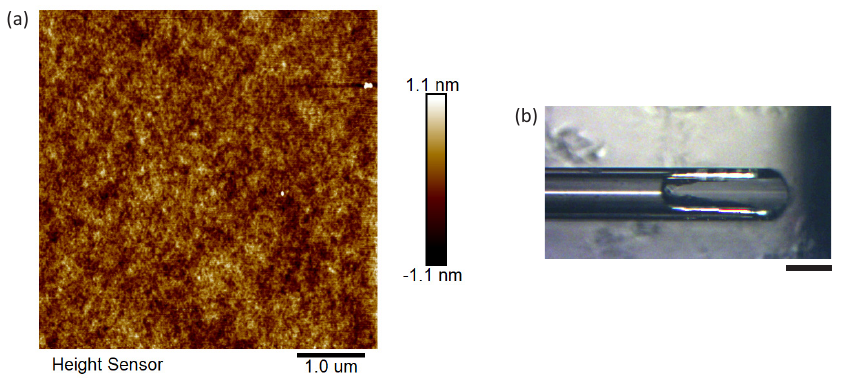}
    \caption{\label{fig:afm}
    (a) Inner surface topography of a glass capillary used in our experiment. (b) Image of a partially fractured capillary. Scale bar is 100~\si{\um}.
  }
  \end{figure}

We repeated the atomic force microscopy (AFM) experiment in Davidson et al. \cite{Davidson2015} to ensure that the inner surface topography of the glass capillary we used does not exhibit any noticeable anisotropy, which may impose any preferred surface orientation.
Using an atomic force microscope (Dimension ICON, Bruker), we scanned the inner surface of a fractured glass capillary (CV1017, Vitrocom, Lot No. 0361C2), as shown in Fig.~\ref{fig:afm}.
The analysis revealed a surface roughness of 0.246~\si{\nm} across 25~\si{\um^2} area, with no discernible texture.

\section{Landau-de Gennes simulation results of the DT configuration under the axial magnetic field}
  \begin{figure}[ht]
  \centering
  \includegraphics{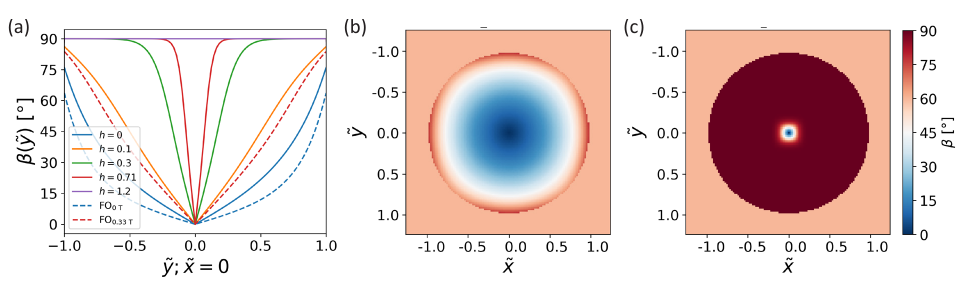}
    \caption{\label{fig:LdGplot}
LdG calculation results with and without the axial magnetic field. (a) The twist angle profiles ($\beta$) along the radial axis according to the calculation method and the magnetic field strength. The profiles from the FO calculations at $B=0$ and $B=0.33\si{\tesla}$ are plotted in dashed lines, respectively. The LdG-simulated director field at the equilibrium state $(-\nabla F < 1~\si{\pico\N})$ at (b) $h=0$ and (c) $h=0.71$, where $F$ is the elastic free energy and $h$ is the parameter concerning the magnetic field explained in the text.
  }
  \end{figure}

We confirm that our results from the Frank-Oseen (FO) free energy approach are semi-quantitatively consistent with the Landau-de Gennes (LdG) calculations. LdG numerical calculations to find the energy-minimizing tensor field $\mathit{Q}$ were performed using the open-Qmin software package \cite{openQmin} with parameters derived from existing chromonic studies \cite{Long2021, Ettinger2022, Ziga2024}. We set the simulation lattice spacing to 1~\si{\um} based on these established parameters. Table \ref{tab:LdGparams} details these parameters. The simulated system's lattice spacing remained larger than the nematic correlation length reported in Zhou et al.~\cite{Zhou2017Cryo}. Our simulations were conducted on a cubic lattice with dimensions $(L_x,L_y,L_z) = (126, 126, 500)$. As Fig.~\ref{fig:LdGplot}(a) illustrates, the LdG calculations reproduced the concave-to-convex transition upon the increase in the field strength $h$, where nondimensionalized magnetic field $\mathbf{\tilde{H}}=\mathbf{H}\sqrt{\mu_0/|A|}=h\hat{z}$. Using our simulation parameters, the $\beta$ profiles from two independent calculations with the same field strength are compared: the profile from the LdG calculation at $h=0.71$ and the one from the FO calculation at $B = 0.33~\si{\tesla}$. However, at high field strength $h>1.2$, all directors align to have $\beta = 90\si{\degree}$, which fails to reproduce the symmetry-breaking instability observed.
  \begin{table}[hb!]
      \centering
      \begin{tabular}{|c|c|c|c|c|c|c|c|c|}
          \hline
          $\tilde{A}$&
          $\tilde{B}$&
          $\tilde{C}$&
          $\tilde{L}_1$&
          $\tilde{L}_2$&
          $\tilde{L}_3$&
          $\tilde{L}_4$&
          $\tilde{L}_6$&
          $\tilde{W}_{\mathrm{FG}}$\\
          \hline
          \hline
          -1&
          -12.326&
          10.058&
          0.0485349&
          1.45605&
          1.35898&
          0&
          0&
          0.68\\
          \hline
      \end{tabular}
      \caption{
      All symbols with tilde annotations are dimensionless. $\tilde{A}$, $\tilde{B}$, and $\tilde{C}$ are normalized by the Landau coefficient $A=16.125\ \si{\joule\per\meter\cubed}$. The elastic constants $L_i$ can be converted to Frank-Oseen constants $K_i$. We set $K_1=K_3=30~\si{pN}$, $K_2=1~\si{pN}$, $K_{24}=15~\si{pN}$, and $\Delta x = 1~\si{\um}$ to satisfy the criterion $\Delta x \ge 1.5~\xi_N$, preventing defect-pinning \cite{Ravnik2009LdG}. The correlation length $\xi_N$ corresponds to $\sqrt{L/(A+BS_0+9CS_0^2/2)}$. Throughout the simulation, the elastic constants $\tilde{L}_i$ are non-dimensionalized by $|A|\Delta x^2$. These parameters yield an equilibrium order parameter of $S_0=0.53$, where $S_0 = (-B+\sqrt{B^2-24AC})/(6C)$.
      }
      \label{tab:LdGparams}
  \end{table}

\section{Estimation of relaxation}
  \begin{figure}[ht]
  \centering
  \includegraphics{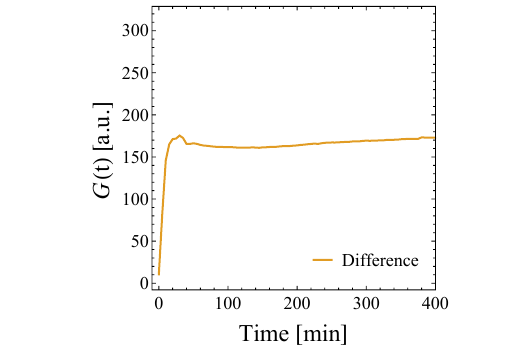}
    \caption{\label{fig:kld}
A representative plot showing that the intensity difference defined by Eq.~\ref{eq:diff} reached a saturation value.
  }
  \end{figure}

To assess if the system reaches its steady state after applying the magnetic field, we estimated the intensity difference $G(t)$ and checked if the curve reached a plateau, as shown in Fig.~\ref{fig:kld}. 
\begin{equation}
    {G(t)=}
    \label{eq:diff}\sum_{i,j}\left|I_{ij}(t_0)-I_{ij}(t)\right|,
\end{equation}
where $I_{ij}(t)$ represent the pixel intensity at the pixel $(i,~j)$ and time $t$, and $t_0$ is the initial time when we started the measurement.
Based on this, we ensured that our sample was fully relaxed after 12~hours of magnetic field application.

\section{Hysteresis in the director configuration}
  \begin{figure}[ht]
  \centering
    \includegraphics{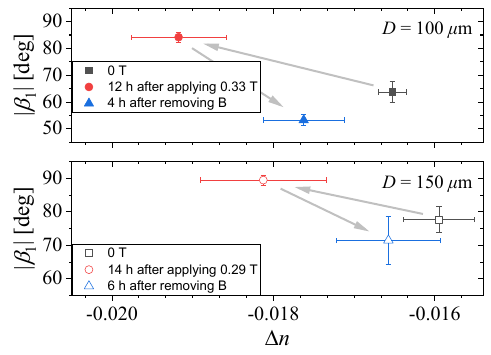}
    \caption{\label{fig:sequence} 
Hysteresis in the twist angle at the capillary wall $\beta_1$ and the birefringence $\Delta n$. We estimate $\beta_1$ and $\Delta n$ of fully relaxed samples at each condition for both 100-\si{\um} diameter (filled symbols) and 150-\si{\um} diameter (hollow symbols) capillaries.
  }
  \end{figure}

We observed hysteresis in the director configuration, where the $\beta$ profile after removing the applied magnetic field did not return to its original state before field application.
As shown in Fig.~\ref{fig:sequence}, regardless of the capillary size $D$, $\beta_1$ increased upon the field application as expected. Note that the degree of birefringence also increased, which presumably indicates changes in LCLC's microstructures. To our interest, despite the full relaxation and our effort to minimize the change in sample concentration, both $\beta$ and $\Delta n$ were not recovered after the magnetic field was removed.

\section{Sunset yellow's DT configurations under an axial magnetic field}
  \begin{figure}[ht]
    \centering
    \includegraphics{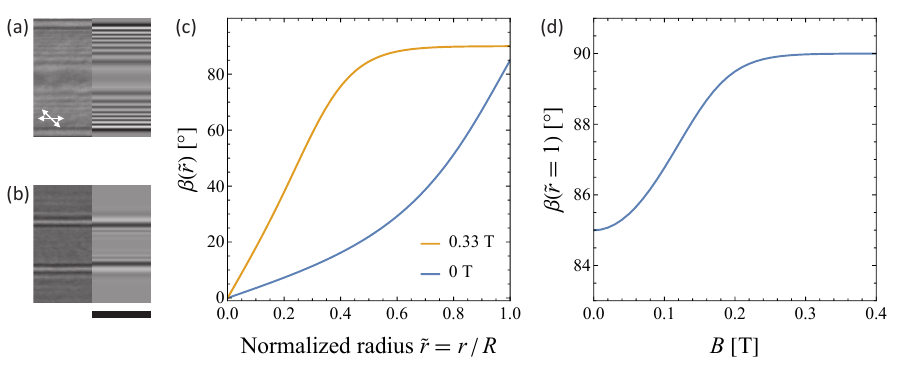}
    \caption{\label{fig:pom-leftover} 
    Experimental and theoretical investigation of SSY's DT configuration under an axial magnetic field. (a and b) POM images of 30~wt\% SSY in 100-\si{\um} diameter capillaries. (a) and (b) display POM images (left) and corresponding Jones calculus-simulated optical textures (right) at $B=0$ and $B=0.33~\si{\tesla}$, respectively. (c) The $\beta(\tilde{r})$ profiles of 30~wt\% SSY at $B=0$ and $B=0.33~\si{\tesla}$. (d) The $\beta(\tilde{r}=1)$ according to the magnetic field strength.
  }
  \end{figure}

We also investigated the magnetic response of Sunset Yellow (SSY), another widely studied LCLC. 
Figure~\ref{fig:pom-leftover}(a) and (b) show POM images before and after the magnetic field application, comparing experimental images (left) with Jones-calculus simulations (right). 
The theoretically calculated $\beta(\tilde{r})$ profile in Fig.~\ref{fig:pom-leftover}(c) indicates that SSY also exhibits the concave-to-convex transition as the field strength increases. For SSY, we used $k_2=1/10$ and $k_{24}=4.6$ \cite{Davidson2015}. Notably, SSY's magnetic anisotropy $\chi_a=-7.2\times10^{-7}$ \cite{Zhou2012PRL} larger than DSCG's $\chi_a=-3.5\times10^{-7}$ makes $\beta_1$ reach the saturated $90\si{\degree}$ at lower $B$ than DSCG. (See Fig.~2(c))